\documentclass{aa}
\usepackage{psfig}
\usepackage{natbib}
\bibpunct{(}{)}{,}{a}{}{,} 
\usepackage{graphicx}

\def\la{\;
\raise0.3ex\hbox{$<$\kern-0.75em\raise-1.1ex\hbox{$\sim$}}\; }
\def\ga{\;
\raise0.3ex\hbox{$>$\kern-0.75em\raise-1.1ex\hbox{$\sim$}}\; }

\newcommand{\dmm}{$\Delta\mu/\mu\,$}
\newcommand{\daa}{$\Delta\alpha/\alpha\,$}
\newcommand{\dFF}{$\Delta F/F\,$}
\newcommand{\kms}{km~s$^{-1}\,$}

\newcommand{\ms}{m~s$^{-1}\,$}

\newcommand{\cmm}{cm$^{-3}\,$}
\newcommand{\etal}{{et al.}}

\begin{document}

\title{An upper limit to the variation in the fundamental constants  
at redshift $z = 5.2$
}
\subtitle{}
\author{S. A. Levshakov\inst{1,2}
\and
F. Combes\inst{3}
\and
F. Boone\inst{4}
\and
I. I. Agafonova\inst{1,2}
\and
D. Reimers\inst{1}
\and
M. G. Kozlov\inst{1,5,6}
}
\institute{
Hamburger Sternwarte, Universit\"at Hamburg,
Gojenbergsweg 112, D-21029 Hamburg, Germany
\and
Ioffe Physical-Technical Institute, 
Polytekhnicheskaya Str. 26, 194021 St.~Petersburg, Russia\\
\email{lev@astro.ioffe.rssi.ru} 
\and
Observatoire de Paris, LERMA, CNRS, 61 Av. de l'Observatoire,
75014 Paris, France
\and
Universit´e de Toulouse, UPS-OMP, CNRS, IRAP, 9 Av. colonel
Roche, BP 44346, 31028, Toulouse Cedex 4, France
\and
Petersburg Nuclear Physics Institute, 188300 Gatchina, Russia
\and
St.~Petersburg Electrotechnical University ``LETI'', Prof. Popov Str. 5, 197376 St.~Petersburg, Russia
}
\date{Received 00  ; Accepted 00}
\abstract
{}
{We constrain a hypothetical variation in the fundamental physical constants 
over the course of cosmic time. 
} 
{We use unique observations of the  CO(7-6) rotational line 
and the [\ion{C}{i}] $^3P_2 - ^3P_1$ fine-structure line towards
a lensed galaxy at redshift $z = 5.2$ to constrain
temporal variations in the constant $F = \alpha^2/\mu$, where $\mu$ is
the electron-to-proton mass ratio and $\alpha$ is the fine-structure constant.
The relative change in $F$ between $z = 0$ and $z = 5.2$,
\dFF $= (F_{\rm obs} - F_{\rm lab})/F_{\rm lab}$,
is estimated from the radial velocity offset, 
$\Delta V = V_{\rm rot} - V_{\rm fs}$,
between the rotational transitions in carbon monoxide
and the fine-structure transition in atomic carbon. 
}
{We find a conservative value $\Delta V = (1 \pm 5)$ \kms\ ($1\sigma$ C.L.), 
which when interpreted in terms of \dFF\ gives \dFF\ $ <2\times10^{-5}$. 
Independent methods restrict the $\mu$-variations 
at the level of $\Delta \mu/\mu < 1\times10^{-7}$ at $z = 0.7$
(look-back time $t_{z0.7} = 6.4$ Gyr).
Assuming that temporal variations in $\mu$, if any, are linear,
this leads to an upper limit on \dmm\ $< 2\times10^{-7}$ at $z = 5.2$
($t_{z5.2} = 12.9$ Gyr). From both constraints on \dFF\ and \dmm,
one obtains for the relative change in $\alpha$ the estimate 
\daa\ $< 8\times10^{-6}$, which 
is at present the tightest limit on \daa\ at 
early cosmological epochs. 
}
{}
\authorrunning{S. A. Levshakov \etal}
\titlerunning{An upper limit to the variation in $\alpha^2/\mu$ at $z = 5.2$}

\maketitle

\section{Introduction}
\label{sect-1}

Space-time variations n the dimensionless physical constants such as
the fine-structure constant, $\alpha = e^2/(\hbar c)$, and
the electron-to-proton mass ratio, $\mu = m_{\rm e}/m_{\rm p}$,
are predicted within the framework of grand unification theories, 
multidimensional theories, and
scalar field theories (for a review, see Uzan 2011).
These constants mediate the strengths of fundamental forces. 
In essence, $\alpha$ is  the coupling constant of the electromagnetic interaction
and $m_{\rm e}$ is  related to the vacuum expectation value 
of the Higgs field, i.e., to the scale of
the weak nuclear force, whereas $m_{\rm p}$ is proportional to the quantum chromodynamic scale, 
$\Lambda_{\rm QCD}$. Hence, $\mu$ probes the ratio of the weak to strong forces of nature.
If detected, a variation in any of these constants would be a manifestation of 
the need to develop new theories
since the Standard Model of particle physics
does not predict any variations in fundamental constants.

Experimentally, the variations in the dimensionless constants  
can be measured by analyzing the relative positions of atomic and molecular
transitions measured in both laboratory and astronomical objects. At present,
the most accurate laboratory results  
on temporal $\alpha$- and $\mu$-variations are  
$\dot{\alpha}/\alpha = (−1.6 \pm 2.3)\times10^{-17}$ yr$^{-1}$
(Rosenband \etal\ 2008) and
$\dot{\mu}/\mu = (1.6 \pm 1.7)\times10^{-15}$ yr$^{-1}$  
(Blatt \etal\ 2008).
For values of $\alpha(t)$ and $\mu(t)$ that are linearly dependent on cosmic time $t$, 
these limits correspond to constraints at redshift $z \sim 2$ 
(look-back time $t_{z2} \sim 10$ Gyr) of
$|\Delta \alpha/\alpha| < 0.4$ ppm (1ppm = $10^{-6}$)
and 
$|\Delta \mu/\mu| < 30$ ppm,
where $\Delta \alpha/\alpha$ (or $\Delta \mu/\mu$)
is a fractional change in $\alpha$ between a reference value $\alpha_1$
and a value $\alpha_2$ measured in a cosmic object given by
$\Delta \alpha/\alpha = (\alpha_2 - \alpha_1)/\alpha_1$.

The most accurate astronomical measurements of the cosmological $\mu$-variation 
are performed in the radio range, which constrain \dmm\ at the level of 0.3 ppm
at $z = 0.89$ (Ellingsen \etal\ 2012) and  0.1 ppm at $z = 0.69$
(Kanekar 2011). The lowest spectroscopic limits on $\alpha$ are
$|\Delta\alpha/\alpha| < 3$ ppm at $z = 0.77$ obtained from
radio observations (Kanekar \etal\ 2012) and
the same order of magnitude 
$|\Delta\alpha/\alpha| \la 3$ ppm at $z \sim 2-3$ derived 
from optical spectra of quasars (Agafonova \etal\ 2011; Molaro \etal\ 2008;
Srianand \etal\ 2007; Quast \etal\ 2004).
To complement these results, the measurements of Murphy \etal\ (2003) 
claim detection of an $\alpha$-variation in about $140$ absorption systems
from quasar spectra observed with the Keck telescope \daa\ $=-5.4\pm1.2$ ppm
in the range $0.2 < z < 3.7$. 
However, Griest \etal\ (2010) showed that the wavelength calibration of these
Keck spectra was affected by systematic and unexplained errors of $\sim 500$ \ms,
which transformed into an error of $\sigma_{\Delta\alpha/\alpha} \sim 30$ ppm. 
The Keck result was not confirmed by the analysis of the VLT spectra of quasars 
acquired from the
southern hemisphere by Webb \etal\ (2011), who found a positive signal of
\daa\ $= 6.1\pm2.0$ ppm at $z \ga 1.8$ in about $150$ systems,
and proposed a model of a cosmic dipole to reconcile their Keck and VLT findings.
A tentative detection of the change in $\alpha$ of \daa\ $= -3.1\pm1.2$ ppm 
was also reported by Kanekar \etal\ (2010) based on observations of 
OH 18~cm lines at $z = 0.25$.

At earlier cosmological epochs $z > 4$, only a handful of spectral measurements 
have been able to provide direct constraints on the temporal variations in $\alpha$ and $\mu$.
Some time ago, we suggested a few methods to probe these variations
at high redshifts by using far-infrared, sub-mm, and
mm transitions in atoms, ions, and molecules  usually observed in
Galactic and extragalactic sources 
(Kozlov \etal\ 2008; Levshakov \etal\ 2008; Levshakov \etal\ 2010a). 
The advantage of fine-structure (FS) transitions is that 
they have the sensitivity to an $\alpha$-variation that is about 30 times
higher than that of UV transitions employed in optical spectroscopy
(Levshakov \etal\ 2008, hereafter L08):
if \daa\ $\neq 0$, then the frequency shift 
$\Delta\omega/\omega_{\rm fs} \equiv (\omega_{z,{\rm fs}} - \omega_{\rm fs})/\omega_{\rm fs}
= 2\Delta\alpha/\alpha$.
To take advantage of the high sensitivity of the FS transitions
in differential measurements of $\alpha$, we suggested to take as 
a reference the low lying rotational lines of carbon monoxide CO since
the rotational frequencies of light molecules are independent of $\alpha$ 
but sensitive to the value of $\mu$, $\Delta\omega/\omega_{\rm rot} = \Delta\mu/\mu$.
In case of variations in $\alpha$ and/or $\mu$ over cosmic time, one should observe
the apparent redshifts of the FS and rotational lines differ of
$\Delta z/(1+z) \equiv \Delta V/c =  \Delta F/F$, where $F = \alpha^2/\mu$,
$\Delta V$ is the difference of the apparent radial velocities, and $c$ is the speed of light (L08). 
The implementation of this method has resulted in 
\dFF~$< 100$ ppm at $z = 6.42$, and \dFF~$< 150$ ppm 
at $z = 4.69$ (L08), and 
\dFF~$<  85$ ppm over the redshift interval $z = 2.3-4.1$ (Curran \etal\ 2011). 

In the present Letter, we show that the [\ion{C}{i}]/CO emission discovered 
by Combes \etal\ (2012, hereafter CRR)
towards the lensed galaxy \object{HLSJ091828.6+514223} at $z = 5.2$ 
allows us to significantly improve this limit.
We note that the look-back time is $t_{z5.2} = 12.9$ Gyr 
for the cosmological parameters 
$H_0 : \Omega_b : \Omega_c : \Omega_\Lambda = 70.5 : 0.046 : 0.228 : 0.726$ 
(Hinshaw \etal\ 2009), which is about $94\%$ 
of the total age of the Universe.

\section{Observations and results}
\label{sect-2}

Details on observations of the $z = 5.2429$ lensed galaxy 
\object{HLSJ091828.6+514223} are given in CRR.
Here we concentrate on the  CO(7-6)/[\ion{C}{i}](2-1) data
since both lines 
were observed simultaneously with the same 2mm receiver 
at the IRAM-interferometer (PdBI).
The spectra were reduced with
the channel width of $\Delta_{\rm ch} = 46.4$ \kms, and the resulting noise rms 
was approximately 0.95 mJy per channel. 

\begin{figure}
\vspace{0.0cm}
\hspace{-0.5cm}\psfig{figure=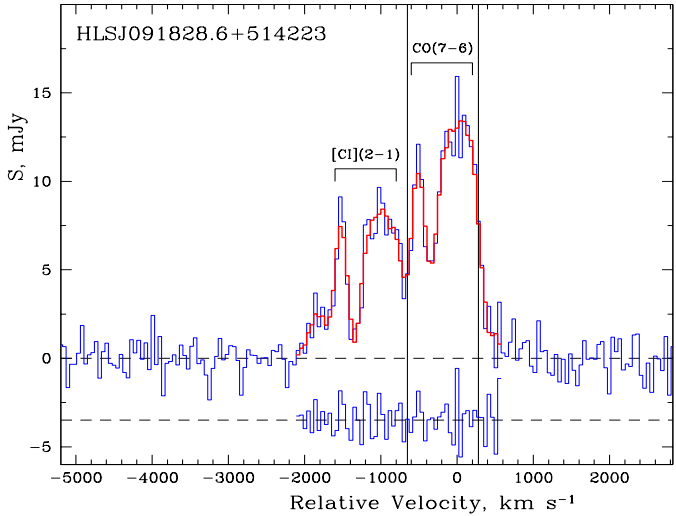,height=10cm,width=11.0cm}
\vspace{-3.5cm}
\caption[]{The CO(7-6) and [\ion{C}{i}](2-1) emission lines (blue color histogram)
towards the lensed galaxy \object{HLSJ091828.6+514223}. 
The velocity scale is given relative to $z=5.2429$. The red
component of the CO(7-6) line is centered at zero velocity.
Its blue component is seen at $V = -510$ \kms.
The red and blue components of [\ion{C}{i}](2-1) are shifted 
with respect to CO(7-6) at $-999.8$ \kms\ owing to their different
rest-frame frequencies. 
Two vertical lines indicate the window used in the bootstrap calculations.
The residuals between the original and smoothed data (red color histogram) are
shown at the bottom. 
}
\label{fg1}
\end{figure}

The observed spectra of CO(J = 7-6) and [\ion{C}{i}](J = 2-1) are shown in Fig.~\ref{fg1}.
We clearly detect
two subcomponents separated by $\simeq 510$ \kms\ and provide in 
Table~2 in CRR the parameters for  
a wider and stronger `red' component of linewidth (FWHM) $w_r \simeq 540$ \kms, 
and a narrow and weaker `blue' component of $w_b \simeq 150$ \kms.
We note that the CO(7-6) and
[\ion{C}{i}](2-1) lines, thanks to their 
close rest-frame frequencies (806651.806 MHz and 809341.97 MHz, respectively),
were observed within one bandwidth, thus their relative positions 
are free from possible instrumental systematics. 

In CRR, the profiles of [\ion{C}{i}](2-1) and CO(7-6) were fitted by two simple Gaussians
(for the blue and red components)
centered at $V_b = -530 \pm 9$ \kms\ and $20 \pm 14$ \kms\ for the former line and at
$V_r = -510 \pm 30$ \kms\ and $4 \pm 10$ \kms\ for the  latter (Table~2 in CRR), 
i.e., the velocity offset between them
is $\Delta V \equiv V_{\rm rot} - V_{\rm fs} = 
20 \pm 31$ \kms\ (blue) and $-16 \pm 17$ \kms\ (red). 
We note that the precision of a single Gaussian line center is
$\sigma = 0.69 \cdot rms \cdot \sqrt{\Delta_{\rm ch} \cdot w }/S_0$
(Landman \etal\ 1982),
where $S_0$ is the peak intensity and $w$ the linewidth. 
For the parameters $ w \sim 150$ \kms, 
$\Delta_{\rm ch} \sim 46$ \kms\, rms $\sim 1$ mJy, and $S_0 \sim 10$ mJy (cf. Fig.~\ref{fg1}),
one finds $\sigma \sim 6$ \kms, which is slightly smaller than the error $\sigma_{V_b}$
quoted above. 

In the present study, we employ a different~-- model-free~-- method 
to measure $\Delta V$. We adopt our approach because
despite their complexity, the CO(7-6) and [\ion{C}{i}](2-1) lines have 
almost identical shapes 
(the scaling factor between their intensities is $\approx 1.6$, as indicated in Table~2 in CRR).
The procedure (``sliding distance'')  is realized in the following way:
one of the chosen profiles (e.g., $T_1$ for CO) is fixed and 
the other ($T_2$ for [\ion{C}{i}]) is shifted relative to $T_1$ sequentially
in small steps, $\delta v$, 
within the velocity interval $[V_1, V_2]$.
At each step, a value of $\Theta^2$~--
the sum of squares of the intensity differences over $n$ points within the line profile~--
is to be calculated ${\Theta}^2(\delta v) = \sum^n_{j=1}[T_{1,j} - T_{2,j}(\delta v)]^2$, 
where $\Theta$ is a kind of distance between the curves, and 
${\Theta}^2$ depends parabolically on $\delta v$ that has its minimum value
when the profiles are aligned. 
The velocity at this minimum is taken as the
offset, $\Delta V$, between the two profiles. 

\begin{figure}
\vspace{0.0cm}
\hspace{-0.5cm}\psfig{figure=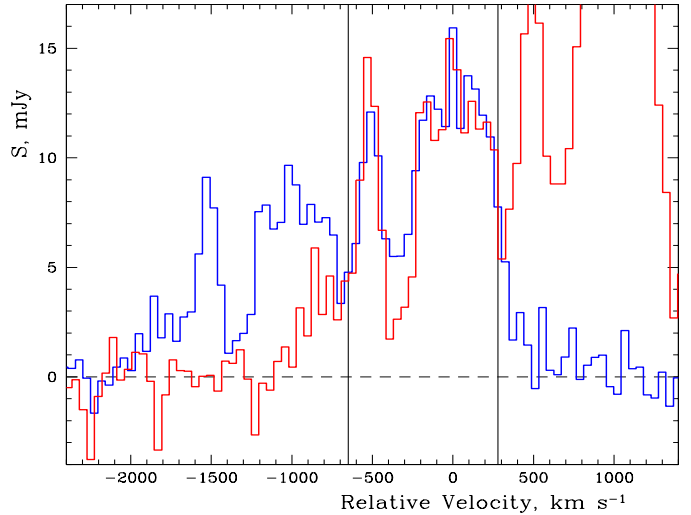,height=10cm,width=11.0cm}
\vspace{-3.5cm}
\caption[]{Same as Fig.~\ref{fg1}, but the original spectrum 
is shifted to +999.8 \kms\ (red color histogram)
and its intensity is zoomed by 
a factor 1.6 to be comparable with the CO(7-6) spectrum
within the window marked by
two vertical lines. This window is used in the bootstrap calculations.
}
\label{fg2}
\end{figure}

In the rest frame, the Doppler velocity shift between the frequencies of [\ion{C}{i}](2-1) 
and CO(7-6) is $\Delta V_0 = -999.8$ \kms.  From Fig.~\ref{fg1}, it can be seen that there is a weak
component at $V \simeq -1800$ \kms\ in the [\ion{C}{i}](2-1) profile\footnote{This negative
velocity wing in [\ion{C}{i}](2-1) could be related to the gas outflows 
(galactic winds) often seen in ultra-luminous infrared galaxies (CRR).}
that may be present in 
the CO(7-6) profile as well, thus distorting part of the [\ion{C}{i}](2-1) profile
around  $V \simeq -800$ \kms. To eliminate the influence of this blending on 
the minimum of $\Theta^2$,  the intensity differences were calculated
only in the velocity window $-650 < V < 280$ \kms\ (marked by the vertical lines in Fig.~\ref{fg1}). 
In addition, to ensure approximately the same intensities in both profiles, 
the intensity of [\ion{C}{i}](2-1) was multiplied by 1.6 (Fig.~\ref{fg2}).

\begin{figure}
\vspace{0.0cm}
\hspace{-1.5cm}\psfig{figure=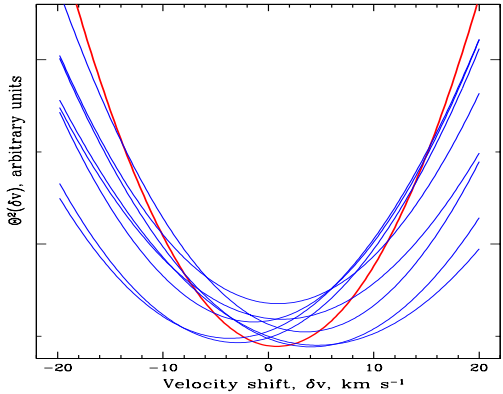,height=7cm,width=10.0cm}
\vspace{-1.0cm}
\caption[]{Examples of the $\Theta^2(\delta v)$ curves 
for the CO(7-6) and [\ion{C}{i}] profiles 
shown in Fig.~\ref{fg2}. The curves were generated by a bootstrap method as described 
in Sect.~\ref{sect-2}. The red line shows $\Theta^2(\delta v)$ of the original data set.
}
\label{fg3}
\end{figure}

\begin{figure}
\vspace{0.0cm}
\hspace{-1.5cm}\psfig{figure=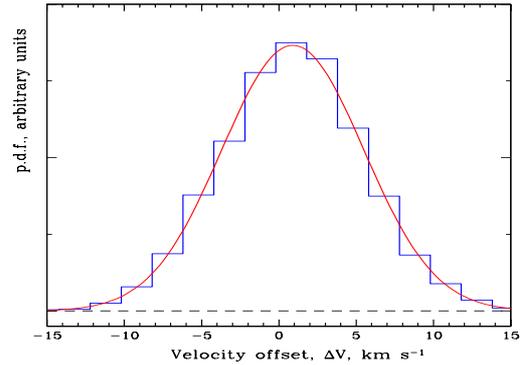,height=7cm,width=10.0cm}
\vspace{-1.5cm}
\caption[]{Bootstrap histogram (probability density function, p.d.f.)
of the velocity offset $\Delta V$ between
the CO(7-6) and [\ion{C}{i}](2-1) profiles shown in Fig.~\ref{fg2}.
Gaussian fit corresponds to $\Delta V = 0.8 \pm 4.6$ \kms\
($1\sigma$ C.L.).
}
\label{fg4}
\end{figure}

The function $\Theta^2$ takes a minimum value at $\Delta V = 0.8$ \kms\ 
(red line in Fig.~\ref{fg3}),  
i.e., the [\ion{C}{i}](2-1) line is shifted
by $-0.8$ \kms\ relative to CO(7-6). 
The error in this offset can be estimated by a bootstrap method, where
artificial data samples are created based on
the statistical properties of the reference data set. In the present context,
this means that we have to create spectral profiles that are statistically
equivalent to the observed ones.
In Fig.~\ref{fg1}, we find both blue- and redward of the emission profiles the extended parts of
the noise spectrum. The statistical analysis of these parts of the spectrum shows
that the noise fluctuations are purely Gaussian (i.e., uncorrelated, with zero skewness and excess
kurtosis) with a dispersion of 0.95 mJy. 
In a subsequent step, we obtain
an emission-line template, i.e., a noise-free emission-line spectrum. Since the fitting
of the Gaussian profiles may be uncertain because of the above-mentioned blending,
the template was prepared by filtering the original data. 
We used a symmetrical Savitzky-Golay filter, 
which does not introduce any 
shifts into the filtered data (Savitzky \& Golay 1964).
The filtered profile is shown by the red color histogram  in Fig.~\ref{fg1}. 
The size of the filtering window
($n = 7$) was chosen from the condition that the dispersion in the residuals
(differences between observed and filtered data) in the velocity range of the observed emission
should be equal to the dispersion in the noise in the emission-free regions (Fig.~\ref{fg1}).
Artificial emission spectra (1000-10000 realizations) were generated by adding the noise
fluctuations taken from the normal distribution
of zero mean and dispersion of 0.95 mJy to the template and 
the minimization of $\Theta^2$ was performed as described above. 
Examples of the $\Theta^2(\delta v)$ curves  for single
realizations are shown in Fig.~\ref{fg3}. The resulting distribution of the 
velocity offsets $\Delta V$ between
the CO(7-6) and [\ion{C}{i}](2-1) profiles is plotted in Fig.~\ref{fg4}. 
It is purely normal with the mean of 0.8 \kms\ and dispersion of 4.6 \kms. 

In calculations, different windows (indicated by vertical lines in 
Fig.~\ref{fg1}) were used to estimate the influence of the possible blending in the
pair [\ion{C}{i}]/CO on the value of the velocity shift. The
dispersion of the $\Delta V$ distribution was found to be very stable,
varying by only a few percent,
whereas the mean of the distribution slightly shifted from the above value of 0.8 \kms\ 
when $[V_1, V_2] = [-650, 280]$ \kms\ to
0.07 \kms\ at $[V_1, V_2] = [-650, 0]$ \kms.
Thus, as a conservative estimate 
the value of $\Delta V = 1 \pm 5$ \kms\ (1$\sigma$ C.L.) 
can be taken for the velocity offset between the
CO(7-6) and [\ion{C}{i}](2-1) lines.
If one interprets this offset in terms of \dFF, then \dFF\ $< 17$ ppm. 
Taking into account that \dmm\ $< 0.1$ ppm at $t_{z0.7} = 6.4$ Gyr and that
the linearly extrapolated limit at $t_{z5.2} = 12.9$ Gyr, \dmm\ $< 0.2$ ppm,
is two orders of magnitude lower than the limit on \dFF, one finds
a constraint on the $\alpha$-variation at $z = 5.2$ of \daa $ < 8$ ppm.

\section{Discussion and conclusions}
\label{sect-3}

By minimizing $\Theta^2$, we have estimated the velocity offset $\Delta V$ 
between the CO(7-6) and [\ion{C}{i}](2-1) profiles
with the precision of 5 \kms, which has provided  
an upper limit to the $F$ variability at $z = 5.2$ that is an order of magnitude
lower then the limit found in a previous survey of the redshifted pairs
of [\ion{C}{i}]/CO lines between $z = 2.3$ and 4.1 (Curran \etal\ 2011).
For comparison, in the Milky Way the [\ion{C}{i}]/CO pairs restrict
\dFF\ at the level of 
$|\Delta F/F| < 0.4$ ppm (Levshakov \etal\ 2010a)
leading to a limit on 
$|\Delta \alpha/\alpha| < 0.2$ ppm since the spatial
variations in $\mu$ are restricted to 
$|\Delta \mu/\mu| < 0.03$ ppm in the disk of the Milky Way
(Levshakov \etal\ 2010b; Levshakov \etal\ 2010c; 
Levshakov \etal\ 2011; Ellingsen \etal\ 2011).

The uncertainty of 5 \kms\ is a statistical error determined by 
the properties of the current observational data, such as 
the S/N ratio, spectral resolution, and spectral line-intensity gradients. 
However, there may be systematic errors that contribute to the total error budget.
In the present case, two potential sources of systematic errors are apparent: 
($i$) velocity shifts caused by observational conditions,  and ($ii$)
velocity shifts due to non-co-spatial distributions of the compared species.
The first source is eliminated thanks to the unique
characteristics of the emission pair [\ion{C}{i}]/CO 
of the $z = 5.2$ galaxy~-- narrow linewidths of the subcomponents and close
rest frame-frequencies~-- which allow us to observe them with the same receiver and 
within the same frequency band.
In this way, we can avoid systematic uncertainties in
the velocity scale calibration during long exposure times.

We consider possible kinematic segregation between CO and C$^0$.
In the Milky Way and in nearby galaxies, 
the C$^0$ emission is closely associated with that
of CO (for references, see Levshakov \etal\ 2010a; Curran \etal\ 2011). 
The carbon-bearing
species C$^0$, C$^+$, and CO are observed in photodissociation regions
(PDRs)~-- neutral regions where chemistry and heating are
regulated by the far-UV photons (Hollenbach \& Tielens 1999).
Photons of energy higher than 11.1 eV
dissociate CO into atomic carbon and oxygen. Since the C$^0$ ionization
potential of 11.3 eV is quite close to the CO dissociation
energy, neutral carbon can be quickly ionized. This suggests that there is a
chemical stratification of the PDR between the lines C$^+$/C$^0$/CO with
increasing depth from the surface of the PDR.
The observed correlation between the
spatial distributions of C$^0$ and CO can be explained by 
clumpy PDR models 
(e.g., Meixner \& Tielens 1995; Spaans  \& van Dishoeck 1997; 
Papadopoulos \etal\ 2004).
In the observed $z = 5.2$ galaxy, the derived excitation temperature 
of the higher-J CO lines, $T_{\rm ex} = 40$K,
and the gas density $n_{{\rm H}_2} = 3.5\times10^3$ \cmm\ (CRR)
just corresponds to the critical density of $n_{\rm cr} \approx 3\times10^3$ \cmm\
required to excite collisionally the J=2 level of the ground state triplet
of C$^0$ (e.g., Levshakov \etal\ 2010a). 
Thus, we can assume that in the present case the
distributions of CO and C$^0$ closely trace each other. 
At $z = 5.2$, we also observe integrated  
emission over all molecular gas-clouds within the galaxy. This naturally
leads to an averaging of the
random fluctuations of the line-of-sight velocity components
$V_{\rm CO}$ and $V_{\rm [CI]}$ and a suppression of the input 
produced by possible deviations
from the co-spatial distribution of the compared species and 
the velocity shift between them.

The uncertainties in the rest-frame frequencies of the analyzed transitions
can be neglected since they are much smaller than the error in the velocity
offset, $\sigma_{\Delta V} = 5$ \kms\ (Klein \etal\ 1998; M\"uller \etal\ 2005).

Thus, we conclude that the limit obtained on the variability of \dFF\ 
at $z = 5.2$ is robust and free from significant systematic errors. 
The error of 5 \kms\ in the velocity offset between CO(7-6) and [\ion{C}{i}](2-1)  
is dominated by the noise fluctuations 
and can be reduced if observations of higher S/N are performed. 
However, even the present constraint on \daa\ $\equiv \Delta V/2c = 2\pm8$ ppm
casts doubts on the dipole model of Webb \etal\ (2011), which predicts
\daa = $-12\pm3$ ppm for the galaxy \object{HLSJ091828.6+514223}.

\begin{acknowledgements}
S.A.L., I.I.A., and M.G.K. are supported by
DFG Sonderforschungsbereich SFB 676 Teilprojekt C4,
and by the RFBR grant 11-02-12284-ofi-m-2011.
\end{acknowledgements}

\end{document}